\documentclass[fleqn, 14pt]{article}
\usepackage{amssymb,amsmath}
\usepackage{geometry}
\newcommand{\be}{\begin{equation}}
\newcommand{\ee}{\end{equation}}
\newcommand{\bel}[1]{\be\label{GrindEQ__#1_}}
\title{EQUATIONS OF QUANTUM AND CLASSICAL MECHANICS AS CONDITIONS OF QUATERNION ALGEBRA STABILITY}
\author{Alexander P. Yefremov\footnote{Address: Russia, 117198 Moscow, Miklukho-Maklaya, str. 6; e-mail address:  \underline{a.yefremov@rudn.ru}.}\\[0.5em]
Institute of Gravitation and Cosmology of \\
Peoples' Friendship University of Russia, Moscow\\[-1.5em]}

\date{}
\begin{document}
\maketitle

\begin{abstract}
Analysis of fundamental structures underlying algebra of quaternion numbers leads to equations equivalent to those of quantum and classical mechanics. A short description of quaternion algebra is given, its units represented in terms of spinors forming vector basis on a complex-number valued 2D space (spinor-plane). Demand of the algebra stability under rotations followed by conformal deformations of the spinor-plane yields a differential equation that in physical space at micro-scale becomes the Schr\"odinger equation. At macro-scale the equation acquires the Hamilton-Jacobi form, and a geometric interpretation of mechanical `minimal action principal' is given. In the presence of a vector field in 3D quaternion space the stability condition takes the form of the Pauli equation for a charged quantum-mechanical particle in magnetic field. \\[1.5em]

\end{abstract}

\section{Introduction}

The mathematics of quaternion (Q-) numbers is known not only as a tool helpful for formulation of conventional equations of physics, but as well as a medium containing many correlations resembling fragments of physical laws. Among them description of non-relativistic and relativistic frames of reference [1, 2], vacuum equations of electromagnetic field [3], expression for Hamiltonian of the Pauli equation [4], geometric appearance of the Yang-Mills intensity tensor as curvature of a Q-space with non-metricity [5]. These fragments, apparently disconnected, none-the-less may be thought of as a manifestation of deep liaison between the last associative division algebra of Q-numbers and the most natural mathematical description of physical phenomena. The study suggested below supports the idea revealing new aspects of this observation. Section 2 contains a review of Q-algebra and its complex-number extension with an accent on the underlying spinor structure that emerge in the form of a specific ``pre-geometric'' space (spinor-plane). In Section 3 a new type of transformation of Q-units, rotations accompanied by conformal deformation of the spinor-plane, are regarded, and equations providing stability of the Q-algebra are derived. Separation of the equation into an imaginary (phase) and real (density) parts leads to a geometric analogue of ``extremum of action'' principal. In Section 4 it is shown that in 3D physical space at the micro-scale the ``stability condition'' is equivalent to the Schr\"odinger equation; at the laboratory macro-scale the phase part takes the form of Hamilton-Jacobi equation, while the real part describes distribution and evolution of the ``semi-mass'' density; a series of solutions is exposed. In Section 5 a vector field is introduced into the ``stability condition'' regarded in the 3D Q-space; this leads to the Schr\"odinger equation comprising the Pauli term. Discussion in Section 6 concludes the study.

\section{Quaternions and pre-geometric spinor-plane}

A quaternion number is an object of the form 
\bel{1}
Q\equiv a\cdot I+b_{k}  {\bf q}_{k},
\ee
$a,\, \, b_{k} $ are real numbers, \textit{I} is a scalar unit, $ {\bf q}_{k} =( {\bf q}_{1} ,\, {\bf q}_{2} ,\, {\bf q}_{3} )$ are three imaginary vector units obeying associative but no more communicative multiplication rule (small Latin indices are 3D, summation rule holds; 
$\delta _{kn},\, \, \varepsilon_{knm}$ are Kronecker and Levi-Civita symbols)
\bel{2}
I\,{\bf q}_{k} = {\bf q}_{k} \, I=\,  {\bf q}_{k} , \quad  {\bf q}_{k}\, {\bf q}_{n} =-\delta _{kn} +\varepsilon _{knm}  {\bf q}_{m}.
\ee
The Q-units frequently are represented by $2\times 2$-matrices; in this case $I$ is the unit matrix, ${\bf q}_{k}$ are Pauli matrices with factor $-i$\textit{ }(scalar imaginary unit). If coefficients $a,\, \, b_{k} $ in Eq.~\eqref{GrindEQ__1_} are complex numbers then the object $Q$ belongs to the algebra of bi-quaternion (BQ-) numbers. The BQ-algebra comprises zero dividers, though ``bad'' mathematically, it fits for physics successfully describing relativistic correlations and massless particles [6]. As well the simplest BQ-numbers ${\bf p}_{k} \equiv i {\bf q}_{k} $ are used in non-relativistic physics as Pauli matrices, forming the multiplication table
\bel{3}
{\bf p}_{k}\, {\bf p}_{n} =\delta _{kn} +i\varepsilon _{knm} {\bf p}_{m}
\ee
with the real scalar part equal to 3D Euclidean metric of  physical space. Q-units $\{ I,\,  {\bf q}_{k} \} $ forming the basis both for Q-numbers and BQ-numbers possess two important properties. First, generalized rotations of vector units as $SL(2, \mathbb{C})$ transformations of the type
\bel{4}
q'_{k} =U {\bf q}_{k} U^{-1}, \quad U\in SL(2, \mathbb{C}), 
\ee
keep the multiplication table \eqref{GrindEQ__2_} invariant. Second, it is proved [7] that each $2\times 2$-matrix-vector Q-unit has a set of right (2D column $\psi^{\pm}$) and left (2D row $\varphi^{\pm}$) eigenvectors with respective eigenvalues $\pm i$, and with the orthogonality and normalization conditions satisfied
\bel{5}
\varphi ^{\mp } \psi ^{\pm } =0, \quad \varphi ^{\pm } \psi ^{\pm } =1.
\ee
If $\varphi ^{\pm } $, $\psi ^{\pm } $ are eigenvectors of $ {\bf q}_{3} $, then the Q-units are compositions of these elementary objects
\bel{6}
I=\psi ^{+} \varphi ^{+} +\psi ^{-} \varphi ^{-},\quad
{\bf q}_{1} =-i(\psi ^{+} \varphi ^{-} +\psi ^{-} \varphi ^{+} ),\quad
{\bf q}_{2} =\psi ^{+} \varphi ^{-} -\psi ^{-} \varphi ^{+},\quad
{\bf q}_{3} =i\, (\psi ^{+} \varphi ^{+} -\psi ^{-} \varphi ^{-} ).
\ee
Eqs.~\eqref{GrindEQ__4_} and \eqref{GrindEQ__6_} mean that $\varphi ^{\pm } $, $\psi ^{\pm } $ are $SL(2, \mathbb{C})$-spinors transformed as $\bar{\psi }^{\pm } =U\psi ^{\pm } $, $\bar{\varphi }^{\pm } =\varphi ^{\pm } \, U^{-1} $, while Eqs.~\eqref{GrindEQ__5_} state that the couple $\psi ^{\pm } $ forms a vector basis in a 2D space with the metric $g=\varphi ^{+} \varphi ^{+} +\varphi ^{-} \varphi ^{-} $, the couple $\varphi ^{\pm } $ being a reciprocal basis of co-vectors [8]. The 2D space named the ``spinor-plane'' is said locally forming ``pre-geometry'' [9] whose each dimension directed by $\psi ^{+} $ or $\psi ^{-} $, due to Eqs.~\eqref{GrindEQ__6_}, is a ``square root'' from 3D dimensions pointed by $ {\bf q}_{k} $, and the only one set of pre-geometric spinor vectors $\varphi ^{\pm } $, $\psi ^{\pm } $ (defining the only one spinor-plane) is sufficient for building a 3D world.

\section{Q-algebra stability under rotation and conformal deformation of spinor-plane}

Generalized $SL(2, \mathbb{C})$ [or 1:2 isomorphic $SO(3,C)$] rotations are believed to exhaust the set of Q-unit-vector's transformations leaving basic rule \eqref{GrindEQ__2_} of Q-algebra intact. But discovery of the spinor-plane widens range of options. 

\subsection{Rotation and conformal deformation }

Consider an ordinary $SU(2)\subset SL(2, \mathbb{C})$ rotation of an initial set of spinors $\psi ^{\pm } ,\, \varphi ^{\pm } $, eigenvectors of $ {\bf q}_{3} $, at real angle $\alpha $ about Q-vector $ {\bf q}_{3} $ (respective Q-triad then rotates at the angle $2\alpha$)
\bel{7}
\bar{\psi }^{\pm } \equiv \left(\cos \alpha +\sin \alpha \,  {\bf q}_{3} \right)\, \psi ^{\pm },\quad
\bar{\varphi }^{\pm } \equiv \left(\cos \alpha -\sin \alpha \,  {\bf q}_{3} \right)\, \varphi ^{\pm },
\ee
followed by a conformal deformation $\sigma \ne 0$ (real) of the spinor-plane
\bel{8}
\Psi ^{\pm } \equiv \sigma \, \bar{\psi }^{\pm } =\sigma \left(\cos \alpha +\sin \alpha {\bf q}_{3} \right)\, \psi ^{\pm },\quad \Phi ^{\pm } \equiv \sigma  \bar{\varphi }^{\pm } =\sigma \left(\cos \alpha -\sin \alpha {\bf q}_{3} \right) \varphi ^{\pm }.
\ee
Since $ {\bf q}_{3} \psi ^{\pm } =\pm i\, \psi ^{\pm } $, $\varphi ^{\pm }  {\bf q}_{3} =\pm i\, \varphi ^{\pm } $ the spinors \eqref{GrindEQ__7_} are $\bar{\psi }^{\pm } =\exp \left(\pm i\alpha \right)\!\psi ^{\pm } $,  $\bar{\varphi }^{\pm } =\exp \left(\mp i\alpha \right)\! \varphi ^{\pm } $; this means that while 3D vector $ {\bf q}_{3}$ (rotation axis) is not changed by the transformation \eqref{GrindEQ__7_}, the spinor-plane changes because its spinors (2D basis) acquire a phase shift. But all objects built of spinors \eqref{GrindEQ__7_} according to the rules \eqref{GrindEQ__6_} are still Q-units. Denote the complex-number scale factor as
\bel{9}
\lambda \equiv \sigma \exp \left(i\, \alpha \right)
\ee
then the spinors \eqref{GrindEQ__8_} rewritten as
\[\Psi ^{+} =\lambda\psi ^{+}, \quad
\Phi ^{+} =\lambda^*\varphi ^{-};\quad
\Psi ^{-} =\lambda^*\psi ^{-},\quad
\Phi ^{-} =\lambda \, \varphi ^{-} \] 
are no more a good material for building Q-units due to presence of the real factor $\sigma \ne 1$. But there is a way to overcome the obstacle.

\subsection{General stability condition}

Let the scale factor \eqref{GrindEQ__9_} be a semi-compact function $\lambda(\eta,\, \xi_{N})$ of a parameter $\eta $ and coordinates $\xi_{N}$ of an abstract $M$-dimensional Euclidean space $P$ ($\eta,\,\xi_{N}$ are unitless), so that the following integral over the space volume $V_{M}$ has finite (normalized) value
\bel{10}
f\equiv \int\limits_{V_{M} }\lambda \lambda^* \, dV_{M} =1.
\ee
Then it is possible to define a good set of Q-units according to the rules \eqref{GrindEQ__6_}
\bel{11}
\!\!\!\!\!\!\!\!\bar{I}=f(\bar{\psi }^{+} \bar{\varphi }^{+} +\bar{\psi }^{-} \bar{\varphi }^{-} )=I,\quad \bar{\bf q}_{3} =if (\bar{\psi }^{+} \bar{\varphi }^{+} -\bar{\psi }^{-} \bar{\varphi }^{-} )= {\bf q}_{3},
\ee
\[\bar{\bf q}_{1} =-if(\bar{\psi }^{+} \bar{\varphi }^{-} +\bar{\psi }^{-} \bar{\varphi }^{+} )=\cos 2\alpha {\bf q}_{1} +\sin 2\alpha {\bf q}_{2},\;\; \bar{\bf q}_{2} =f (\bar{\psi }^{+} \bar{\varphi }^{-} +\bar{\psi }^{-} \bar{\varphi }^{+} )=-\sin 2\alpha {\bf q}_{1} +\cos 2\alpha{\bf q}_{2}.\] 

``Exteriorly'' the units \eqref{GrindEQ__11_} coincide with those built from the spinor \eqref{GrindEQ__7_}, i.e. the 3D Q-triad is rotated about ${\bf q}_{3}$ at angle 2$\alpha$; but this set of Q-units has an ``interior'' dependence on the functional $f$. Demand that Eq.~\eqref{GrindEQ__10_} be explicitly independent of the parameter $\eta$: $\partial_\eta f=0$; this requirement evokes the continuity equation
\bel{12}
\partial _{\eta } (\lambda \lambda^*)+\nabla (\lambda \lambda^*\vec{k})=0,\ee
$\partial _{\eta } \equiv \partial {\rm /}\partial \eta $, $\nabla \equiv \partial {\rm /}\partial \xi _{N} $. Let vector $\vec{k}$, ``propagation'' of the Q-triad \eqref{GrindEQ__11_} in the space $P$, indicate maximal increase of the phase
\bel{13}
\vec{k}\equiv \nabla \alpha =\frac{i}{2} \nabla \left(\ln \frac{\lambda^*}{\lambda } \right)=\frac{i}{2} \left(\frac{1}{\lambda^*} \nabla \lambda^*-\frac{1}{\lambda } \nabla \lambda \right).
\ee
Then substitution of the definition \eqref{GrindEQ__13_} into Eq.~\eqref{GrindEQ__12_} after a simple algebra yields 
\bel{14}
\partial _{\eta } \lambda -\frac{i}{2} \left(\nabla ^{2} \lambda -2 W\lambda \right)+\left[\partial _{\eta } \lambda^*+\frac{i}{2} \left(\nabla ^{2} \lambda^*-2 W\lambda^*\right)\right]\exp (2 i\alpha )=0,
\ee
the free term $iW\lambda \lambda^*$ is added and subtracted, $W(\eta,\, \xi_{N})$ is a (unitless) real function. If Eq.~\eqref{GrindEQ__14_} is valid for any phase value, then its mutually conjugated parts vanish separately, in particular 
\bel{15}
\left[\partial _{\eta } -\frac{i}{2} \, \left(\nabla ^{2} -2\, W\right)\right]\lambda =0.
\ee
Eq.~\eqref{GrindEQ__15_} is a general condition of the Q-algebra stability under transformations \eqref{GrindEQ__7_}, \eqref{GrindEQ__8_} of the spinor-plane over the space $P$, or it can be regarded as a projection of 3D Eq.~\eqref{GrindEQ__12_} onto the 2D spinor-plane; in any case it is a pure mathematical equation. But one notes that the form of Eq.~\eqref{GrindEQ__15_} resembles that of the equation of non-relativistic quantum mechanics.

\subsection{Split equations}

Substitution of the function $\lambda $ in the form \eqref{GrindEQ__9_} splits Eq.~\eqref{GrindEQ__15_} into a real part, equation for the ``semi-length'' density $\sigma $
\bel{16}
\partial _{\eta } \sigma +\nabla \sigma \, \nabla \alpha +\frac{1}{2} \sigma \, \nabla ^{2} \alpha =0,
\ee
that is a ``Q-spinor square root'' of the Eq.~\eqref{GrindEQ__12_}, and an imaginary part, equation for the phase
\bel{17} \partial _{\eta } \alpha +\frac{1}{2} \, \left(\nabla \alpha \right)\, ^{2} +\, W'\, =0 \ee
where  
\bel{18}
W'\equiv W-\frac{\nabla ^{2} \sigma }{2\sigma }. \ee
One notes that Eq.~\eqref{GrindEQ__17_} has the form of the Hamilton-Jacobi equation. 

\subsection{Minimal phase}

In 3D space $P$ the spinors \eqref{GrindEQ__8_} have minimal phase value (or an arc scribed by the end of e.g. vector $\bar{\bf q}_{1} $of the propagating and rotating Q-triad has minimal length) if the propagation vector $\vec{k}$ is always aligned with the rotation axis $\bar{\bf q}_{3}$. The phase value on a finite segment of trajectory can be found from Eq.~\eqref{GrindEQ__17_} and expression of the phase's full derivative $d_{\eta } \alpha =\partial _{\eta } \alpha +\, \partial _{\eta } \vec{\xi } \nabla \alpha $ 
\bel{19}
\alpha =\int\limits_{\eta _{1} }^{\eta _{2} }\left[\partial _{\eta } \vec{\xi }\, \nabla \alpha -\frac{1}{2} \, \left(\nabla \alpha \right)\, ^{2} -W'\right] \, d\eta.
\ee
Vanishing variation of the functional \eqref{GrindEQ__19_} provides its minimum and the Euler-Lagrange-type equations, while subintegral expression generalizes the Lagrange function of classical mechanics. Alternatively one can demand that cosine of the angle between $\vec{k}$ and $\bar{\bf q}_{3}$ be always maximal on all possible paths of the Q-triad initial point. 

\section{Physical applications}

Derived above equations \eqref{GrindEQ__15_} and \eqref{GrindEQ__17_} would stay just a mathematical event if similar expressions were not earlier established adequately describing physical phenomena. Therefore casting these equations into known physical forms seems expedient; this also may be helpful for better understanding the physical facts themselves. Below the parameter $\eta$ and the space $P$ are associated respectively with universal (Newtonian) time and the 3D physical space $\xi_{N} \to \xi_{k} $; so the unitless coordinates should acquire physical units $x_{k} \sim \varepsilon\xi _{k} $, $t\sim \tau\eta $; as well transit from mathematical to physical equations dictates setting length and time scales.

\subsection{Micro-scale: quantum mechanics}

Let the conformal spinor-factor ``semi-length'' density $\sigma $ physically mean endowing the spin-surface with a ``semi-mass'' density $[\sigma ]=g^{1/2}~cm^{-3/2}$, so that the vector Q-triad acquires a characteristic (small) mass $m=\int\limits_{V_{3} }\sigma ^{2} dV_{3}  $ determined in a 3D space volume $V_{3} $ with characteristic (small) length $\varepsilon $; speed of the processes at this length is chosen to be compared with the fundamental velocity \textit{c}, characteristic (small) interval of time is $\tau \sim \varepsilon /c$. At this scale a 3D particle may be conceived as a distributed in space but compact mass with ``frozen in'' rotating Q-triad, the phase $\alpha $ changing with time. One can introduce its characteristic proper angular momentum $m\varepsilon ^{2} /\tau \sim mc\,\varepsilon $, the Planck constant is fitting in units and value experimental magnitude $mc\,\varepsilon \sim \hbar $. Thus a short length-and-time micro ($\mu$-) scale is established $\varepsilon \sim \frac{\hbar }{m\, c} $, $\tau \sim \frac{\hbar }{m\, c^{2} } $ (for an electron $\varepsilon \sim 10^{-11}~cm$, $\tau \sim 10^{-21}~s$), the unitless parameter and coordinates become $\eta \sim \frac{mc^{2} }{\hbar } t$, $\xi _{k} \sim \frac{m\, c}{\hbar } x_{k} $, and their substitution into Eq.~\eqref{GrindEQ__15_} yields precisely the Schr\"odinger equation
\bel{20}
\left(i\hbar \partial _{t} +\frac{\hbar ^{2} }{2m} \nabla ^{2} -V\right)\, \lambda =0
\ee
with the free function $V(t,x_{k})\equiv mc^{2} W$. Probably Eq.~\eqref{GrindEQ__20_} can be satisfied by functions in the form different from that of Eq.~\eqref{GrindEQ__9_}, while Eqs.~\eqref{GrindEQ__16_} and \eqref{GrindEQ__17_}, apparently equivalent to Eq.~\eqref{GrindEQ__15_} [provided Eq.~\eqref{GrindEQ__9_} holds], nonetheless will be used for transit to classical mechanics. 

\subsection{Transition to macro-scale: classical mechanics}

At the $\mu$-scale ($\varepsilon,\, \tau $) in the 3D physical world Eq.~\eqref{GrindEQ__17_} is rewritten as
\bel{21}
\hbar \, \partial _{t} \alpha +\frac{\hbar ^{2} }{2m} \, \left(\nabla \alpha \right)\, ^{2} +\, mc^{2} \, W'\, =0.
\ee
Let a Q-triad move along the laboratory macro (M-) coordinate $x=u(t_{2}-t_{1} )$ with slowly changing velocity $u\sim{\rm const}$, but rotating so that $2\alpha =\Omega(t_{2} -t_{1} )$, the cycle frequency and transverse lengths measured at the $\mu$-scale $\Omega \sim 1/\tau $, $\varepsilon \ll x$; $\vec{k}$ is collinear with $\bar{\bf q}_{3} $. Then the formula for length of a helix arc (of radius $\varepsilon $) on a segment of the trajectory 
\[
l_{0} =\int\limits_{t_{1} }^{t_{2} }u\sqrt{1+\frac{\varepsilon ^{2} \Omega ^{2} }{u^{2} } }  dt\cong x+\frac{\varepsilon ^{2} \, }{u\tau } \alpha =x+\frac{\hbar }{mu} \alpha
\] 
gives expression for the phase 
\bel{22}
\alpha =\frac{mu\, (l_{0} -x)}{\hbar } \equiv \frac{S}{\hbar },\ee
with $S\equiv \hbar \alpha =mu(l_{0}-x)$, $mu=-\partial \, S/\partial x$. 

Substitution of Eq.~\eqref{GrindEQ__22_} into Eq.~\eqref{GrindEQ__21_} with the notation $V_{c} \equiv mc^{2} W'$ gives precisely the classical Hamilton-Jacobi equation for the action $S$
\bel{23}
\partial _{t} S+\frac{1}{2m} \, \left(\nabla S\right)\, ^{2} +\, V_{c} \, =0.
\ee
Eq.~\eqref{GrindEQ__16_} at the $\mu$-scale has the form
\[
\partial _{t} \sigma +\frac{\hbar }{m} \left(\nabla \sigma \, \nabla \alpha +\frac{1}{2} \sigma \, \nabla ^{2} \alpha \right)=0;
\] 
at the M-scale Eq.~\eqref{GrindEQ__22_} holds, so the ``classical'' equation for the ``semi-mass'' density is
\bel{24}
\partial _{t} \sigma +\frac{1}{m} \left(\nabla \sigma \, \nabla S+\frac{1}{2} \sigma \, \nabla ^{2} S\right)=0;
\ee
one expects that Eq.~\eqref{GrindEQ__24_} describes evolution of a massive particle. 

Finely analyze the arbitrary function defined by Eq.~\eqref{GrindEQ__18_}. In Eq.~\eqref{GrindEQ__21_} at the $\mu$-scale it acquires the energy units, so a physical assumption can be made. Eq.~\eqref{GrindEQ__23_} states that at the M-scale $V_{c} $ plays the role of classical potential energy of a mechanical system, but the variable $\sigma $ must be regarded at the scale $\varepsilon $ 
\bel{25}
V_{c} =mc^{2} W-\frac{\hbar ^{2} }{2m} \frac{\nabla ^{2} \sigma }{\sigma }.
\ee
This means that the energy term $mc^{2} W$, apart from $V_{c}$, comprises a $\mu$-scale term $ V_{\mu } $ 
\[ mc^{2} W=V_{c} +V_{\mu }; \] 
$V_{\mu } $ and the last term of Eq.~\eqref{GrindEQ__25_} form the following static equation (with $\kappa \equiv \, 2m/\hbar ^{2} $)
\bel{26}
\nabla ^{2} \sigma -\kappa V_{\mu } \sigma =0.
\ee
Find characteristic solutions of Eq.~\eqref{GrindEQ__26_}, noting in advance that $V_{\mu } \ne 0$ because solutions of the equation $\nabla ^{2} \sigma =0$ make the integral $\int\limits_{\infty} \sigma^{2}dV$ diverge. 

\subsection{Free motion} 

The simplest $\mu$-scale constant energy $V_{\mu } $ is to be proportional to $mc^{2} $. Let the energy term be $\kappa V_{\mu } =4 m^{2} c^{2} /\hbar ^{2} =1/(2\varepsilon )^{2} $, then Eq.~\eqref{GrindEQ__26_} takes the form $\nabla ^{2} \sigma =\sigma /(2\varepsilon )^{2} $ and has (in Cartesian coordinates) the solution
$\sigma =\tilde{\sigma }(t)\exp [-(x+y+z)/(2\varepsilon )]$ well defined on the semi-space volume $x,y,z\in [0,\, +\infty )$. If at the M-scale $V_{c} =0$, then Eq.~\eqref{GrindEQ__23_} describes free motion, e.g. along coordinate $x$, with $(-\partial _{x} S)/m\equiv u{\rm \; }={\rm const}$; Eq.~\eqref{GrindEQ__24_} reduces to $\partial _{t} \sigma =\sigma u/(2\varepsilon )$ and leads to the wave solution $\sigma =\sigma _{0} \exp [(ut-x-y-z)/(2\varepsilon )]$,
($\sigma _{0} ={\rm const}$) and compact mass integral 
\[
\int\limits_{V_{3}^{+} }\sigma ^{2} dV_{3}^{+}  =\sigma _{0}^{2} \varepsilon ^{3} \int\limits_{0}^{\infty }e^{-(x-ut)/\varepsilon } d[(x-ut){\rm /}\varepsilon {\rm ]} \int\limits_{0}^{\infty }e^{-y/\varepsilon } d(y/\varepsilon ) \int\limits_{0}^{\infty }e^{-z/\varepsilon } d(z/\varepsilon )=\sigma _{0}^{2} \varepsilon ^{3} \equiv m
\] 
equivalently represented by a product of three $\delta$-functions.

\subsection{Spherically symmetric (static) solutions}

Let the $\mu$-scale energy term be a (rational) polynomial function of radius $\kappa V_{\mu }(r)=B+C\, r^{\gamma } $ where \textit{B, C } are real constants, $\gamma$ is an integer, then Eq.~\eqref{GrindEQ__26_} in spherical coordinates takes the form
\[
\frac{1}{r^2} \frac{d}{dr} \left(r^{2} \frac{d\sigma }{dr} \right)-\kappa V_\mu\sigma=0
\] 
and has a set of solutions yielding compact mass-integrals in the volume $ r\in [0, +\infty )$:

~~~(i) constant energy ($\gamma =0$), ``Yukawa-density'' solution 
\[\kappa V_{\mu } =B,\quad \sigma =\frac{\tilde{\sigma }(t)}{r} \exp \left(-\sqrt{B} \, r\right),\] 
~~~(ii) constant + ``Newton gravity'' energy ($\gamma =-1$) solution 
\[\kappa V_{\mu } =B-\frac{2\sqrt{B} }{r},\quad \sigma =\tilde{\sigma }(t)\exp \left(-\sqrt{B} \, r\right),\] 
~~~(iii) constant + Hook (oscillator) energy ($\gamma =2$), ``Gauss-density'' solution 
\[\kappa V_{\mu } =-3B+\, (B\, r)^{2}, \quad \sigma =\tilde{\sigma }(t)\exp \left(-\frac{B\, r^{2} }{2} \right),\] 
~~~(iv) constant + Hook energy ($\gamma =2$), ``Gauss-Yukawa-density'' solution 
\[\kappa V_{\mu } =-B+\, (B\, r)^{2} ,\quad \sigma =\frac{\tilde{\sigma }(t)}{r} \exp \left(-\frac{B\, r^{2} }{2} \right),\] 
with the $\mu$-scale constant for all these solutions$\, B\sim 1/\varepsilon ^{2} $. Other functions satisfying Eqs.~\eqref{GrindEQ__23_}, \eqref{GrindEQ__24_} and \eqref{GrindEQ__26_} can be sought for as physical applications. 

\section{Vector field, Q-space and Pauli equation}

Return back to stability conditions of the Q-basis \eqref{GrindEQ__11_} that below will be regarded over 3D space with coordinates $\xi _{k} $ and with a free parameter $\eta $. Let the continuity Eq.~\eqref{GrindEQ__12_} still hold but with the ``propagation'' vector distorted by a vector field 
\bel{27}
k_{n} \equiv \partial _{n} \alpha \, \, +A_{n},
\ee
where $\partial _{n} \equiv \partial /\partial \xi _{n} $. The field $ A_{n} (\xi _{j} ,\eta )$ evokes reminiscences of electrodynamics, and it is the place to remind two facts. First, Maxwell had written his equations in the quaternion form (as the most adequate), and second, Fueter [3] showed that vacuum Maxwell equations are mathematically equivalent to the requirement of differentiability of a Q-vector function of Q-variable. So introduction of a 3D vector field in Eq.~\eqref{GrindEQ__27_} motivates matching it (hence any other 3D vector) to a vector Q-triad, or to regard involved equations over 3D quaternion space [10] with the metric \eqref{GrindEQ__3_} having positive Euclidean part
\bel{28}
g_{kn} \equiv \, {\bf p}_{k} {\bf p}_{n} =\delta _{kn} +i\varepsilon_{knm} {\bf p}_{m}.
\ee
Moreover, recall that the integral \eqref{GrindEQ__10_} describes a spinor normalization condition, so that there should exist an extension of Eq.~\eqref{GrindEQ__15_} valid for a spinor function (e.g. of the positive parity). Therefore Eq.~\eqref{GrindEQ__12_} with the ``propagation'' vector \eqref{GrindEQ__27_} must be written for spinors, so built on the scalar part of the metric \eqref{GrindEQ__28_}
\bel{29}
\partial _{\eta } (\varphi ^{+} \lambda^*\lambda \psi ^{+} )+\frac{1}{2} ({\bf p}_{n} {\bf p}_{m} +{\bf p}_{m} {\bf p}_{n} )\partial _{n} [\varphi ^{+} \lambda^*\lambda \psi ^{+} (\partial _{m} \alpha +A_{m} )]=0.
\ee
Using the procedure similar to that of the Section 3 represent Eq.~\eqref{GrindEQ__29_} identically as
\[
(\lambda^*\varphi ^{+} ) \times \left[\left(\partial _{\eta } -\frac{i}{2} \partial _{k} \partial _{k} +\frac{1}{2} \partial _{k} A_{k} +A_{k} \partial _{k} +\frac{i}{2} A_{k} A_{k} +\frac{i}{2} \varepsilon _{kmj} {\bf p}_{j} \partial _{m} A_{k} +iW\right)(\lambda \psi ^{+} )\right]+\] 
\[+\left[\left(\partial _{\eta } +\frac{i}{2} \partial _{k} \partial _{k} +\frac{1}{2} \partial _{k} A_{k} +A_{k} \partial _{k} -\frac{i}{2} A_{k} A_{k} +\frac{i}{2} \varepsilon _{mkj} {\bf p}_{j} \partial _{m} A_{k} -iW\right)\, (\lambda^*\varphi ^{+} )\right]\times (\lambda \psi ^{+} )=0,\]
a free term $\displaystyle i(\lambda^*\varphi ^{+} )\times \left(\frac{1}{2} A_{k} A_{k} +W\right)\times (\lambda \psi ^{+} )$ is added and subtracted. If the last equation is valid for any phase and for eigenvectors of any Q-unit, then it brakes up into mutually conjugated ``square-root'' parts; in particular equation for a spinor-vector $\Psi ^{+} =\lambda \psi ^{+} $ is
\bel{30}
\left[i\partial _{\eta } -\frac{1}{2} (-i\partial _{k} +A_{k} )^{2} -\frac{1}{2} {\bf p}_{j} B_{j} -W\right]\Psi ^{+} =0,
\ee
with $B_{j} \equiv \varepsilon _{mkj} \partial _{m} A_{k} $. As a physical application at the $\mu$-scale Eq.~\eqref{GrindEQ__30_} becomes precisely the Schr\"odinger-Pauli equation for an electrically charged particle in a magnetic field. The specific Pauli term (at once with Bohr-magneton factor) $\frac{e\hbar }{2mc} {\bf p}_{j} B_{j} $ was earlier found [4] in a heuristic assumption that the Q-space non-symmetric structure interacts with a magnetic vector-potential.

\section{Discussion}

The suggested above derivation of a series of famous physical equations from pure mathematical considerations can hardly be judged a casual coincidence. A mosaic of isolated fragments of physical laws earlier discovered in the Q-math medium, in this study started developing into a distinct picture, at least what concerns equations of non-relativistic quantum and classical mechanics. Nonetheless it is a certain surprise that this picture is revealing itself not in the `3D-space \& time' physical world but on a complex-number valued 2D spinor-plane formed by spinor functions, the simplest (for the moment) elements of quaternion algebra. 

There can be two types of attitude to this pre-geometric spinor-plane. On the one hand it can be thought of just as a mathematically admitted plane-tool virtually (and instantly) scanning the space volume and ``printing'' the data on a ``world screen'' in the form of respective ``square-root'' equations. On the other hand it can be a manifestation of physical reality thus meaning that the 3D space has a hardly conceivable interior structure.

Apart of conventional Schr\"odinger, Pauli and Hamilton-Jacobi equations of mechanics the suggested approach leads to a set of specific Eqs.~\eqref{GrindEQ__16_}, \eqref{GrindEQ__17_} and \eqref{GrindEQ__18_} that can be in more details analyzed for application at small length-and-time scales. As well, natural treatment of the mechanical ``extremum of action'' principle as minimal rotational angle of a propagating Q-triad (or as always maximal on the path cosine of angle between the axis of rotation and the propagation arrow) forces one to again think over the universality of the principal nowadays applied to any physical system. 

Instead the approach offers consideration of different variants of the continuity equation including distortions of the propagation vector by various reasons and possible dependence of the Q-space metric on coordinates and time, as in Einstein's gravity.

So in this study the fine quaternion mathematics, early shown to contain relativity, electrodynamics and Yang-Mills objects, now additionally exposes equations of quantum mechanics. This ``natural symbiosis'' in the Q-math midst may lead to promising unifications.

\end{document}